\documentclass{aa}  
\usepackage{graphicx}
\usepackage{physics,amsmath}
\usepackage[normalem]{ulem}
\usepackage{txfonts}

\usepackage{natbib}
\bibpunct{(}{)}{;}{a}{}{,} 
\usepackage[colorlinks=true, allcolors = blue]{hyperref}

\hypersetup{
  colorlinks   = true,
  urlcolor     = blue,
  linkcolor    = blue,
  citecolor   = blue 
}
\usepackage{xcolor}

\begin{document}

  \title{Suppression of photospheric velocity fluctuations in strongly magnetic O-stars in radiation-magnetohydrodynamic simulations}
    \author{A. ud-Doula
          \inst{1,2}\fnmsep\thanks{corresponding author}
          \and
          J.O. Sundqvist
           \inst{2}
          \and
          N. Narechania
          \inst{3}
          \and
          D. Debnath \inst{2}
          \and
          N. Moens \inst{2}
          \and
          R. Keppens \inst{3}
          }
   \institute{Penn State Scranton, 120 Ridge View Drive, Dunmore, PA 18512, USA\\
              \email{asif@psu.edu}
              \and
   Instituut voor Sterrenkunde, KU Leuven, Celestijnenlaan 200D, 
              3001 Leuven, Belgium,
          \and
         Centre for mathematical Plasma Astrophysics, Department of Mathematics, KU Leuven, Celestijnenlaan 200B, 3001 Leuven, Belgium\\       
             }

   \date{Received; Accepted}

   \titlerunning{Turbulence suppression}
   \authorrunning{A. ud-Doula et al.}

  \abstract
    {O-stars generally show clear signs of strong line-broadening (in addition to rotational broadening) in their photospheric absorption lines (typically referred to as `macroturbulence'), believed to originate in a turbulent sub-surface zone associated with enhanced opacities due to recombination of iron-group elements (at $T \sim 150-200$ kK). O-stars with detected global magnetic fields also display such macroturbulence; the sole exception to this is NGC 1624-2, which also  has the strongest (by far) detected field of the known magnetic O-stars. It has been suggested that this lack of additional line-broadening is because NGC 1624-2's exceptionally strong magnetic field might be able able to suppress the turbulent velocity field generated in the iron opacity peak zone.}
   {Here, we test this hypothesis based on two-dimensional (2D), time-dependent, radiation magnetohydrodynamical (RMHD) box-in-a-star simulations for O-stars that encapsulate the deeper sub-surface atmosphere (down to T $\sim$ 400 kK), the stellar photosphere, and the onset of the supersonic line-driven wind in one unified approach. To study potential suppression of atmospheric velocity fluctuations, we extend our previous non-magnetic O-star radiation-hydrodynamic (RHD) simulations to include magnetic fields of varying strengths and orientations.} 
   {We use \texttt{MPI-AMRVAC} which is a Fortran 90 based publically available parallel finite volume code that is highly modular. We use the recently added RMHD module to perform all our simulations here.}
   {For moderately strong magnetic cases ($\sim$ 1 kG) the simulated atmospheres are highly structured characterised by large root-mean-square velocities, and our results are qualitatively similar to those found in previous non-magnetic studies. By contrast, we find that a strong horizontal magnetic field in excess of 10 kG can indeed suppress the large velocity fluctuations and thus stabilise (and thereby also inflate) the atmosphere of a typical early O-star in the Galaxy. On the other hand, an equally strong radial field is only able to suppress horizontal motions, and as a consequence these models exhibit significant radial fluctuations.}
   {Our simulations provide an overall physical rationale as to why NGC 1624-2 with its strong $\sim$ 20 kG dipolar field lacks the large macroturbulent line broadening that all other known slowly rotating early O-stars exhibit. However, they also highlight the importance of field geometry for controlling the atmospheric dynamics in massive and luminous stars that are strongly magnetic, tentatively suggesting latitudinal dependence of macroturbulence and basic photospheric parameters.}


   \maketitle
   
\section{Introduction}

Optical absorption-line spectroscopy of O-stars has revealed that rotation is not the only macroscopic broadening mechanism at play in their atmospheres  \citep[e.g.,][]{Conti_1977,howarth_1997,simon_2017}. The additional broadening component is typically very broad -- for early O-stars full-width half-maxima (FWHM) are $\sim 50-100$ km/s -- and often have a Gaussian-like shape. When spectra are fitted by means of 1D model atmosphere codes it is included as an ad-hoc very large  macroturbulence \citep{simon_2007,sundqvist_2013, simon_2017}. However, modern multi-dimensional radiation-hydrodynamic (RHD) atmospheric simulations \citep{Jiang_2015, schultz_2022,debnath_2024} 
and earlier theoretical studies \citep[e.g.,][]{Cantiello_2009, Grassitelli_2015} demonstrate that such large turbulent velocity fields naturally occur in the envelopes of luminous O-stars.

In these models, atmospheric structure originates in a radiatively and convectively unstable region located slightly beneath the stellar surface, at temperatures $T \sim 150-200$ kK, associated with a peak in the Rosseland mean opacity due to recombination effects of iron-group elements \citep{Rogers_1992}. Due to the enhanced opacities, local pockets of gas in this sub-surface region receive a radiation force that can locally exceed the inward gravitational pull and be strongly accelerated upwards toward the upper atmosphere, where they further interact with the line-driven outflow \citep{cak1975} that is initiated from just above the variable optical photosphere \citep{debnath_2024}. This complex interplay then creates large photospheric turbulent velocities, typically on order $\varv_{\rm turb} \sim 50 - 100$ km/s for early and luminous O-stars in the Galaxy \citep{Jiang_2015, debnath_2024}. Recent spectral synthesis performed directly from such multi-D models have shown that indeed the typical widths of observed absorption line profiles can be quite well matched \citep{schultz_2022}.  

In parallel with this development, spectropolarimetric observations have revealed that $\sim 7-10$ \% of massive main-sequence stars posses strong magnetic fields \citep{Wade_2012, Grunhut_2017}. The detected fields are large-scale, ordered, often with a dominant dipole component, and stable over long time scales. As such, their origin is not related to any possible dynamo mechanism operating in their atmospheres; instead they are believed to be surviving fossil fields \citep{Jermyn_2020}, which potentially may be related to stellar mergers \citep{Schneider_2019}.
Interestingly, magnetic O-stars are also very slow rotators, which likely is a result of strong magnetic braking in their powerful line-driven wind outflows \citep[e.g.,][]{ud_doula_2009}. 

\citet{sundqvist_2013} 
used the fact that rotational broadening is negligible in magnetic O-stars to empirically investigate the occurrence of turbulent motions in their atmospheres from analysis of their optical absorption lines. They found that the only magnetic O-star that did not possess significant macroscopic line broadening was NGC 1624-2, which also is the O-star with the strongest (by far) detected magnetic field. Specifically, NGC 1624-2 has an inferred dipole magnetic field of $\sim$ 20 kG whereas the other magnetic O-stars have fields $\sim$ 1 kG \citep{Wade_2012b}. Using simple scaling analysis arguments then, it was suggested that the atmosphere of NGC 1624-2 might be stabilised against velocity fluctuations since even at the iron opacity peak it was estimated that magnetic pressure should be higher than gas pressure \citep[see also][]{MacDonald_2019, Jermyn_2020}. On the other hand, for the other magnetic O-stars (with an order of magnitude weaker fields), equipartition should occur already very near the stellar surface, well above the iron opacity bump zone. As such, the fluctuations leading to macroturbulence in these stars would not be suppressed, in agreement with observations of their absorption line profiles. 

This mechanism for magnetic inhibition of turbulent motion would then be quite analogous to that believed to occur in the magnetic and chemically peculiar Ap stars  \citep{Michaud_1970, Landstreet_1998}, with the key difference that for O-stars the turbulence originates in a sub-surface instability zone associated with iron recombination. More specifically,  in Ap stars, the magnetic fields are believed to inhibit convective motion in the atmosphere, leading to chemical stratification \citep{Michaud_1970}. This process allows heavy elements like iron and rare earths to accumulate in localized regions, which manifests in their unique spectral lines. The suppression of turbulence by magnetic fields in the atmosphere is central to the development of these chemical peculiarities. In these stars, the magnetic field disrupts the mixing in their atmospheres, allowing diffusion processes to separate elements and create the observed overabundances. In this context, it may also be worth mentioning that although instabilities originate beneath the surface for O-stars, numerical simulations show that the resulting turbulent zone extend up to the stellar photosphere and into the wind \citep{Jiang_2015, schultz_2022, debnath_2024}. How potential suppression of this turbulence by a strong magnetic field might affect chemical abundance patterns in this regime remain an open question.

\citet{Jiang_2017} computed multi-dimensional RMHD models of massive-star envelopes covering the iron opacity bump region, focusing on field strengths up to 1 kG. They found that such fields were not able to suppress the very large characteristic   turbulent velocities associated with corresponding non-magnetic simulations. In this paper we test if even stronger (up to 20 kG) magnetic fields indeed are able to inhibit atmospheric velocity fluctuations in O-stars based on 2D simplified models that preclude any dynamo effects that can lead to further magnetic field amplifications. We utilize recently developed methods of radiation-hydrodynamic simulations for hot, luminous and massive star atmospheres \citep{moens2022radiation, debnath_2024} and extend them to include initially uniform magnetic field with fixed orientation. In section \ref{methods}, we discuss our numerical methods including initial and boundary conditions. We present simulation results in section \ref{simulation_results} and a general discussion in section \ref{Discussion}, before a short summary is given in section \ref{summary}. 

\section{Simulation methods}
\label{methods}
Based on initial 3D simulations of Wolf-Rayet outflows by \citet{Moens_2022A},
\citet{debnath_2024} conducted time-dependent, two-dimensional (2D) simulations of O-star atmospheres using a flux-limited diffusion (FLD) radiation-hydrodynamics (RHD) finite volume grid in a `box-in-a-star' configuration, including line-driving opacities and corrections for spherical divergence. Building on that work, we extend the analysis by incorporating an initially uniform magnetic field of different strengths threading the atmosphere of a representative early O-star in the Galaxy.

\subsection{Radiation-magnetohydrodynamics} 

All our models were performed using  \texttt{MPI-AMRVAC} which is a Fortran 90 based parallel code often used for magnetohydrodynamic (MHD) simulations of solar and astrophysical plasmas~\citep{keppens2012parallel,porth2014mpi,xia_2018, keppens2023mpi}. 
\texttt{MPI-AMRVAC} is highly modular with options for different numerical schemes and switches to include various physics terms in the governing equations enabling users to integrate newer modules into the existing code with relative ease. Most recently, \cite{moens2022radiation} have implemented a flux-limited diffusion (FLD) approach in \texttt{MPI-AMRVAC} to couple the hydrodynamics module with radiation. For this work, we extend this FLD approach to include equations of magnetohydrodynamics. Details of this numerical scheme can be found in Narechania et al. (submitted) but we briefly summarize some key points for the convenience of the reader.

\texttt{MPI-AMRVAC} numerically solves the RMHD equations on finite volume meshes in conservative form. These equations, consisting of the equations of conservation of mass, momentum, energy and magnetic flux, including the effect of radiation and gravity in the form of source terms, are given by 
\begin{equation}\label{eq:mhd_mass}
\frac{\partial {\rho}}{\partial {t}} + {\nabla} \cdot \left({\rho} {\mbox{\bf v}}\right) = 0,
\end{equation}
\begin{equation}\label{eq:mhd_mom}
\begin{split}
\frac{\partial (\rho{\mbox{\bf v}})}{\partial {t}} + {\nabla}\cdot\left(\rho \mbox{\bf v} \mbox{\bf v} - \mbox{\bf B} \mbox{\bf B} + {\left(p + \frac{\mbox{\bf B} \cdot \mbox{\bf B}}{2}\right)}\mbox{\bf I}\right) = \bf f_r + f_{grav},
\end{split}
\end{equation}
\begin{equation}\label{eq:mhd_energy}
\begin{split}
\frac{\partial e}{\partial t} + 
\nabla\cdot\left(\left(e + p + \frac{\mbox{\bf B}\cdot\mbox{\bf B}}{2}\right)\mbox{\bf v} - (\mbox{\bf B}\cdot \mbox{\bf v})\mbox{\bf B} \right) = \mbox{\bf v}\cdot \left( {\bf f_r+f_{grav}}\right) + \dot{q},
\end{split}
\end{equation}
\begin{equation}\label{eq:mhd_mag}
\frac{\partial{\mbox{\bf B}}}{\partial t} + {\nabla}\cdot \left(\mbox{\bf v} \mbox{\bf B} - \mbox{\bf B} \mbox{\bf v}\right) = \bf 0.
\end{equation}
Here, ${\rho}$ is the plasma density, ${\mbox{\bf v}}$ is the plasma velocity, $p$ is the plasma pressure, ${\mbox{\bf B}}$ is the magnetic field,  {${\mbox{\bf I}}$ is the identity tensor}, ${\bf f_r}$ is the radiation force described below and ${\bf f_{grav}}$ is the gravitational force of a point stellar mass $M_\star$:
\begin{equation}
\mathbf{f}_{\text{grav}} = - \rho \frac{GM_\star}{r^2} \hat{\mathbf{r}}.
\end{equation}
The total energy, $e$, is the sum of the kinetic, internal and magnetic energies and is given by
\begin{equation}\label{eq:energy_density}
e = \frac{p}{\gamma-1} + \frac{\rho v^2}{2} + \frac{B^2}{2}, 
\end{equation}
where $\gamma$ is the adiabatic index of the gas. The ideal gas law given by
\begin{equation}\label{eq:ideal_gas_law}
p = \frac{k_B T_g}{m_p \mu} \rho,
\end{equation}
provides the closure relation between the plasma temperature, pressure and density. Here,  $m_p$ is proton mass, $\mu$ is the mean molecular weight, $T_g$ is the plasma temperature and $k_B$ is the Boltzmann constant. Throughout this paper, we assume constant $\mu = 0.61$ and $\gamma = 5/3$ (see also \citealt{debnath_2024}).

Additionally, the magnetic field must also satisfy the divergence-free condition:
\begin{equation}\label{eq:div_B}
{\nabla} \cdot {\mbox{\bf B}} = 0.
\end{equation}
For the purpose of the current work, the discrete divergence is maintained using the  Powell source term method \citep{powell1994approximate}.

The $\bf f_r$ term seen in the momentum equation is the radiation force written as
\begin{equation}\label{eq:rad_force}
{\bf f_r} = \frac{\rho \kappa_F \mbox{\bf F}}{c},
\end{equation}
where $c$ is the speed of light, $\mbox{\bf F}$ is the radiation flux vector, and $\kappa_F$ is the flux mean opacity.
The $\dot{q}$ term in the energy equation is the term describing the plasma's interaction with radiation through heating and cooling, and is given by 
\begin{equation}\label{eq:heating}
\dot{q} = c \kappa_E \rho E - 4 \kappa_P \rho \sigma T_g^4.
\end{equation}
Here, $E$ is the radiation energy density, $\kappa_P$ is the Planck opacity, $\kappa_E$ is the energy density opacity, and $\sigma$ is the Stefan-Boltzmann constant. Since $\bf f_r$ and $\dot{q}$ are both dependent on the radiation field, we also need a conservation equation for the radiation energy density. This equation is given by the frequency-integrated zeroth angular moment of the radiative transfer equation
\begin{equation}\label{eq:mhd_r_e}
\frac{\partial {E}}{\partial {t}} + {\nabla} \cdot \left({E} {\mbox{\bf v}}\right) + {\nabla} \cdot \mbox{\bf F} + \mbox{\bf P} : {\nabla} \mbox{\bf v}  = -\dot{q}.
\end{equation}
The above equation is written in the co-moving frame of the fluid, due to the ease of computing opacities in this frame, as opposed to an inertial frame of reference. Here, $\mbox{\bf P}$ is the radiation pressure tensor and $\mbox{\bf P} : {\nabla} \mbox{\bf v}$ is the `photon tiring' term.

The FLD-approximation by \citet{levermore1981flux} is used to close the radiation flux vector $\mbox{\bf F}$:
\begin{equation}\label{eq:fld_F}
\mbox{\bf F} = -\frac{c \lambda}{\rho \kappa_F} \nabla E,
\end{equation}
For the flux limiter $\lambda$, we employ the analytic relation
\begin{equation}\label{eq:fld_lambda}
\lambda = \frac{2 + R}{6 + 3R +R^2},
\end{equation}
as proposed by \citet{levermore1981flux}. Here, the dimensionless radiation energy density gradient, $R$, is given by
\begin{equation}\label{eq:fld_R}
R = \frac{|\nabla E|}{\rho \kappa E}.
\end{equation}
This limiter $\lambda$ ensures that the radiation flux does not exceed the physical limit, $cE$, in the optically thin free-streaming limit. On the other hand, for the diffusion limit,  the accurate $\mbox{\bf F} = -c \nabla E/(3\rho \kappa)$ is recovered. 

Finally,  $\mbox{\bf P}$ is evaluated as a function of $E$ as
\begin{equation}\label{eq:fld_P}
\mbox{\bf P} = \mbox{\bf f} E,
\end{equation}
where $\mbox{\bf f}$ is the Eddington tensor as proposed by \citet{turner_2001}. 
The numerical methods used to evaluate all these above terms have been described in detail by \citet{moens2022radiation}. Rigorous testing of the new RMHD module is realized in Narechania et al. (submitted). We use the magnetic field splitting technique in the RMHD module as described in \citet{xia_2018} (and references therein). As in previous work \citep{Moens_2022A, debnath_2024}, we assume flux, energy, and Planck opacities to be equal, and compute opacities following the hybrid formalism by \citet{Poniatowski_2022}. This combines tabulated Rosseland means for deeper sub-surface layers with line-driving opacities effective in optically thin parts of the atmosphere. The so-called finite-disc effect of line-driving \citep{pauldrach_1986} is included following \citet{debnath_2024}. 

\subsection{Numerical grid}
All our simulations presented here use a Cartesian mesh. Although \texttt{MPI-AMRVAC} has the option for adaptive mesh refinement (AMR), we perform all our models using a fixed static grid with 128 in radial and 64 points in horizontal direction. 
For our `box-in-a-star' model of an O-star atmosphere, we use the same stellar parameters as employed by \citet{debnath_2024} for their O4 star model, as shown in Table \ref{table:Models}. Thus, we set the lower boundary to a fixed radius $R_0 = 13.54 R_\odot$, yielding a lower boundary temperature $T_0 \sim 450kK$ deep inside the stellar envelope to model the entire `iron opacity bump' region. Our aim is to study atmospheric velocity fluctuations generated in the stellar envelope. As such, we limit the outer boundary in the radial direction to $r=2R_0$. 
Such a relatively small range also ensures that our assumption of uniform magnetic field within the box is reasonable. The horizontal coordinate covers a range $-0.25 R_0$ to $0.25 R_0$.

\subsection{Initial and boundary conditions}
\label{Initial_conditions}

For our initial and boundary conditions in all simulations presented here, we again closely follow the procedure outlined in \citet{debnath_2024} and refer the interested reader for more details there. 
In brief, we assume a spherically symmetric, stationary, and analytic wind structure approximated by a `$\beta$' velocity law and mass density according to an initial assumed mass-loss rate. An analytic wind is smoothly connected to a lower atmosphere assumed to be in hydrostatic and thermal equilibrium. The magnetic field is  uniform throughout the computational domain. 

Boundary conditions for MHD models often can be challenging. Here, in the horizontal direction, boundaries are assumed to be periodic for all conserved quantities. 
At the lower boundary, the density is held constant as defined by the initial conditions, and the values for the ghost cells are extrapolated assuming hydrostatic equilibrium. The gas momentum is free to vary and is extrapolated from the first active cell to the ghost cells under the assumption of mass conservation. The radiation energy density $E_{rad}$ is extracted from its gradient by using Eq. \ref{eq:fld_F} with a fixed input radiative luminosity at the lower boundary.

At the outer boundary, outflow boundary conditions are applied for all quantities except for the radiation energy $E_{rad}$ which is analytically solved from the outer boundary radius to $r  \rightarrow \infty$ as described in \citet{moens2022radiation} and \citet{debnath_2024}. However, the focus of this work is not on the stellar wind itself, and therefore, further details of the wind outflow are not addressed in this paper.

\subsection{Parameter study}
\label{Parameter_study}

To investigate the minimum magnetic field strength required to suppress turbulence in the subsurface region of an O4 atmosphere, we conducted numerical simulations with uniform magnetic fields of 1 kG, 10 kG, and 20 kG, oriented in both radial and horizontal directions. The  magnetic field configurations were introduced at time $t=0$ in a computational domain that evolved for approximately 9 physical days which is about 20 dynamical timescales if we assume typical macroturbulence velocity of 300 km s$^{-1}$ in the subsurface layer.  By directly comparing these magnetic simulations\footnote{We note a slight inconsistency in our approach: while all hydrodynamical quantities incorporate spherical divergence, the evolution of the magnetic field does not. Given that the range of our computational domain is relatively small, we do not expect our results and conclusions to change qualitatively.} to a non-magnetic control model, we aimed to identify the critical magnetic field threshold necessary to stabilize the atmospheric dynamics.

\section{Simulation results} 
\label{simulation_results}
 
\subsection{Classification of models}
\label{classification}
Our 2D simulations here specify mass, radius and radiative luminosity at the lower boundary. 
Both the photospheric radius and the effective temperature $T_{\rm eff}$, defined at the average photospheric radius $R_\ast$ where the mean optical depth $\tau \approx 2/3$, are emergent properties. Following the methodology of \citet{debnath_2024}, we select a model representative of a typical O4 star.

Keeping this model fixed, we only vary the strength and topology of the uniform magnetic field. We classify our models into three categories based on the strength of the magnetic field: (i) weak  $B \sim 100$G, (ii) moderately strong  $B \sim 1$kG and (iii) strong $B \ge 10$kG. These fields can be either in the radial or horizontal directions. In our local approach, they thus represent magnetic polar (radial) and equatorial (horizontal) viewing angles for an external observer.

In the non-magnetic O4 simulation presented by \citet{debnath_2024} structure starts appearing just below the iron-opacity peak at $\sim$ 200 kK, where the gas becomes subject to convective, radiative, and also Rayleigh-Taylor instabilities (\citealt{bs03}, van der Sijpt et al., in prep.). Local pockets of gas there experience radiative accelerations that exceed gravity, so that they shoot up from these deep layers into the upper and cooler atmosphere. In those upper atmospheric parts, Rosseland mean opacities are lower than in the deeper layers, meaning the plasma is decelerated. While some gas particles then indeed turn over and fall back into the deep atmosphere, others become subject to line-driving opacities and thus are re-accelerated in the region around and just above the variable optical photosphere (launching an average supersonic wind outflow). This complex interplay creates large velocity dispersions ($\ga 100$ km/s for the case studied here) in the photospheric layers of such O-star RHD simulations. These large velocity dispersions provide significant turbulent pressure support, energy transport by enthalpy (=convection) is very inefficient in these luminous O-star atmospheres. For the case discussed here, radiative enthalpy is much higher than gas enthalpy but still only provides $\la 10$ \% of the total energy transport even at the peak of the iron opacity-bump, with the rest of the energy being carried by diffuse radiation. On the other hand, {the large velocity dispersions give rise to a large photospheric turbulent pressure, typically on the same order as the radiation pressure and significantly higher than gas pressure, with associated turbulent velocities $\varv_{\rm turb} = \sqrt{\langle \rho \varv_r^2 \rangle / \langle \rho \rangle}$ that in turn are on the same order as the measured velocity dispersions (see figure 16 in \citealt{debnath_2024}, and corresponding definitions in their Sect. 4.1)}.  

{We note that while the analytic closure relation in our FLD approximation method is physically correct for deep optically thick layers, it is only an approximation in regions where the atmosphere transitions from optically thick to optically thin, which may affect our predicted photospheric velocity dispersions. For those transitional regions, it would be desirable to instead apply a closure relation based on numerical solutions to the radiative transfer equation, as in the Variable Eddington Tensor (VET) models by \citet{Jiang_2017}. On the other hand, those simulations then do not include effects of line-driving, which for O-stars have a very strong impact on the dynamics near and above the stellar surface \citep{puls08, debnath_2024}. Since (because of the complicated frequency-dependence of flux-weighted line opacities) even 1D, stationary atmospheric models accounting properly for such line-driving are very computationally expensive \citep{sander_2017, sundqvist_bjorklund_2019}, it remains a challenge to include this effect in VET methods applied to multi-D simulations. However, since the focus here is on investigating if a strong magnetic field can suppress velocity fluctuations originating in deep optically thick layers, we expect that our overall results should be fairly unaffected by these radiative transfer complications, at least qualitatively. Indeed, this seems to be supported also by the overall qualitative agreement between our simulations and those by \citet{Jiang_2017}. Specifically,} our simulations with weak magnetic fields closely resemble those of our corresponding non-magnetic model, though the former exhibits a slight increase in velocity dispersion in the subsurface layers. This minor enhancement in turbulent velocity fluctuations for massive stars with weak magnetic fields qualitatively aligns with the findings of \cite{Jiang_2017}, who observed similar behaviour in their study. Since the aim here is to examine whether stronger magnetic fields can suppress such atmospheric velocity fluctuations, we focus primarily on discussing our simulation runs with stronger fields in this paper. 

\begin{table*}
\caption{Fundamental parameters for the $\langle 2D \rangle$ O-star model studied in this paper. From left to right, the columns display the model name, effective temperature, stellar mass, radius, luminosity, Eddington ratio, and surface gravity. Angle brackets denote time and horizontally averaged quantities. Note that some of values differ slightly from \citet{debnath_2024} as they used different mesh resolution.}       
\label{table:Models}      
\centering          
\begin{tabular}{c | c c c c c c }  
\hline\hline       
Model & $\left<T_{\rm eff} [kK]\right>$ & $M_\star/M_\odot$ & $ \langle R_\star \rangle/R_\odot$ & $log_{10} \left(\left<L_\star\right>/L_\odot\right)$ & $\left<L_\star\right>/L_{\rm edd}$& $ log_{10} \left<g_\star\right>$ \\  
\hline                    
 
   $\rm{O}4$ & 39.98  & 58.3 & 16.79 & 5.81  & 0.27 & 3.75 \\ 
\hline                  
\end{tabular}
\end{table*}

\subsection{Basic characteristics of magnetic simulations}

\begin{figure*}
   \centering
   \includegraphics[scale=0.70]{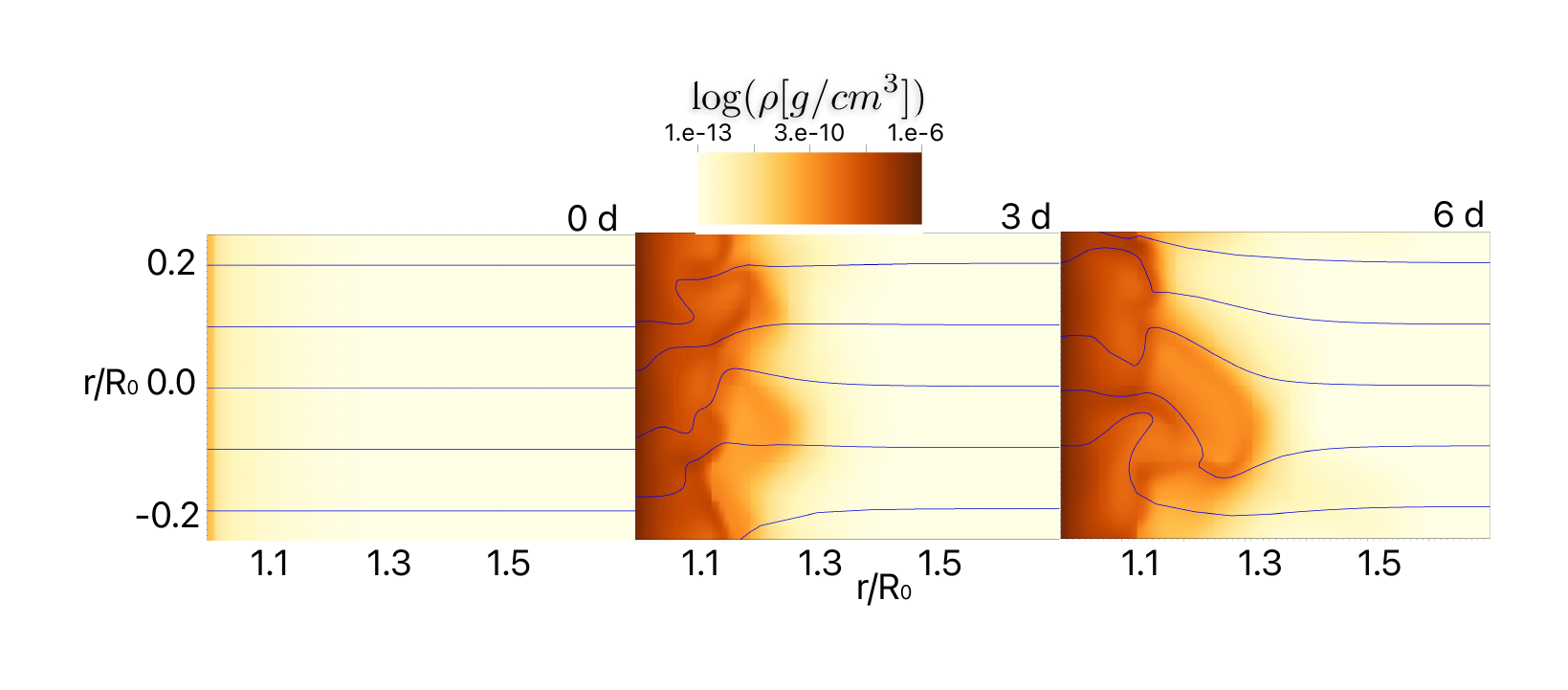}
   \caption{An illustrative temporal evolution of a magnetic model with a 1 kG radial magnetic field. The color scale represents the logarithm of density in {\rm cgs} unit, while the blue lines depict magnetic field lines. Beginning with a spherically symmetric initial condition (left panel) threaded by a uniform magnetic field in radial direction, the model {develops significant density and velocity variations} after 2-4 days (middle panel, time = 3 days) and maintains this until the end of the simulation at approximately 9 days (right panel, time = 6 days).}
              \label{Fig-evol}%
    \end{figure*}

Figure \ref{Fig-evol} shows a typical temporal evolution of our moderately strong magnetic ($B=1$kG) model with the initial magnetic field oriented in the radial direction. The plots show logarithm of density plotted against the spatial coordinates normalized to lower boundary radius spatial coordinates. The blue lines represent the magnetic field. 

After approximately one day of evolution, the model develops {density and velocity fluctuations} in a way similar to the process described above for the non-magnetic simulations. 
As both gas and radiation pressures are higher than the magnetic pressure, clear deviations from the radial topology emerge as the model progresses. In these 1kG simulations, the magnetic field is not sufficiently strong to suppress turbulent motion in the deeper subsurface layers near the iron opacity peak. This is also evident from the significant deformation of magnetic field lines observed in the later stages of evolution (right two panels of figure \ref{Fig-evol}). For weak magnetic cases ($B<100$G), such deformations of magnetic field lines are even more extreme, and the overall evolution closely resembles that of the non-magnetic case. These results are broadly consistent with the 3D RMHD models presented by 
\citet{Jiang_2017}.

\begin{figure*}
   \centering
   \includegraphics[scale=0.60]{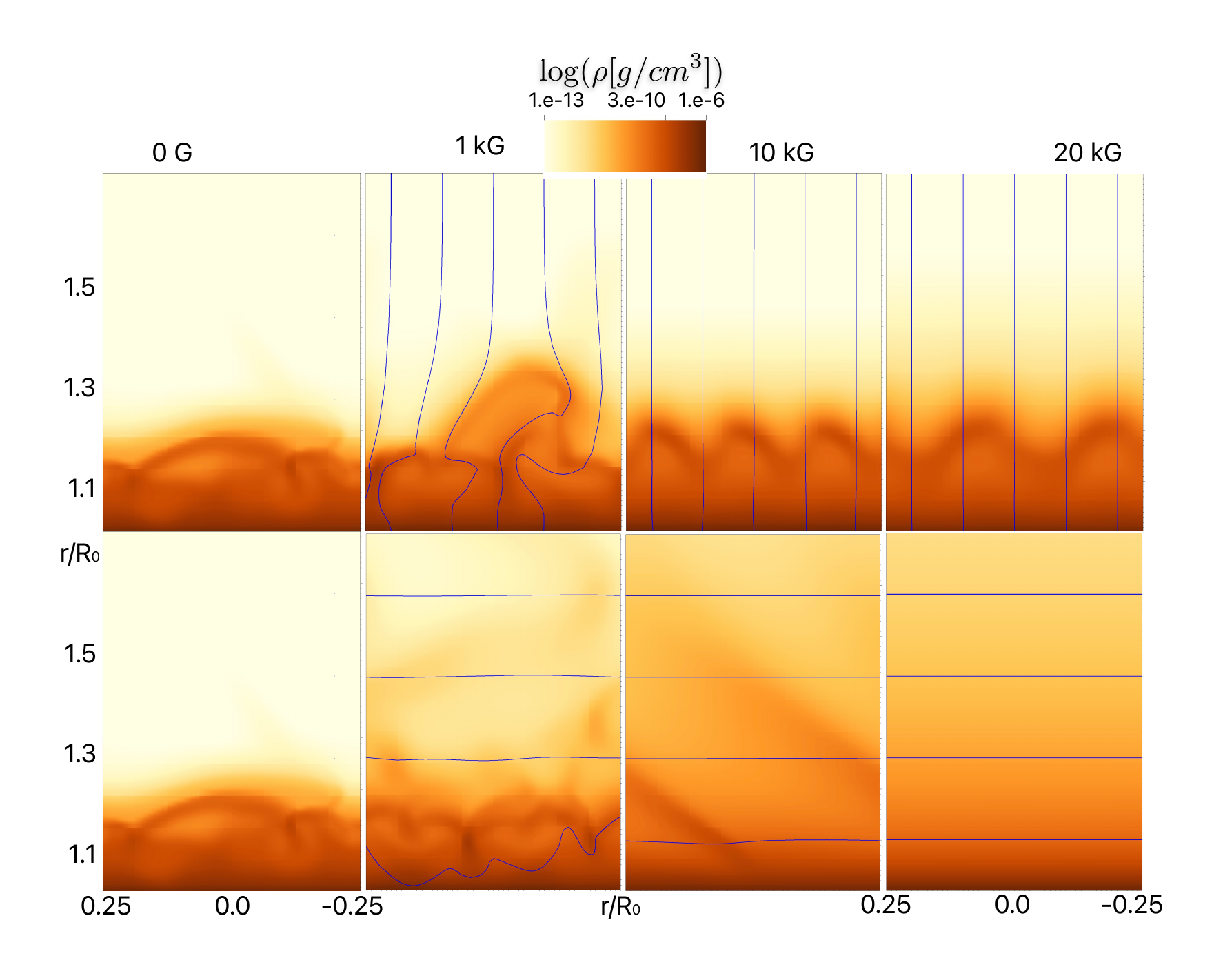}
   \caption{Similar to Fig. \ref{Fig-evol}, a comparison of the late evolutionary states between the non-magnetic case (left column) and various magnetic models. The second column from the right shows models with a 1 kG magnetic field in the radial (top row) and horizontal (bottom row) directions. The third and fourth columns display models with 10 kG and 20 kG magnetic fields, respectively.}
   
              \label{Fig-comparison}%
    \end{figure*}

Figure \ref{Fig-comparison} presents the logarithm of density for the final snapshot of each simulation, taken after approximately 9 days of evolution, for all magnetic field models. By this stage, all simulations have reached a quasi-steady or quasi-periodic state, with no discernible influence from the initial conditions. For a direct comparison, the left column depicts the corresponding non-magnetic case. Progressing to the right, the models feature increasing magnetic field strengths of 1, 10, and 20 kG. The top row displays models with a radial magnetic field, while the bottom row shows those with a horizontal magnetic field orientation.

For the 1 kG models, regardless of whether the magnetic field is oriented radially or horizontally, {strong velocity and density fluctuations} remains a dominant feature throughout the simulation. This also suggests that moderately strong magnetic fields are insufficient to suppress the convective and radiative instabilities driving {this extensive structure development} in the subsurface layers. The field may perturb the flow, but it lacks the strength to significantly alter the gas dynamics and suppress the strong turbulent motions.

In contrast, the behavior changes substantially in the strong magnetic field models. When the magnetic field is aligned radially, it effectively suppresses horizontal gas motions but permits piston-like, oscillatory motions along the radial direction. These motions operate on a timescale of about half a day, indicating a dynamic balance where the magnetic field restricts turbulent flows but allows for periodic compressions and expansions of the gas. This is likely related to the interplay between magnetic pressure and gas pressure in largely confining the flow along the field lines. This also might be the result of the 2D nature of our simulations that artificially imposes symmetry in the third dimension.

For models with strong horizontal magnetic fields, the suppression is even more pronounced. Here, the magnetic field restricts both radial and horizontal motions, effectively quenching {significant atmospheric structure from developing}. The horizontal orientation of the field here clearly inhibits the vertical gas motions that drive large velocity fluctuations, leading to a more stable envelope that more resembles a radiative and strongly magnetic, quasi-hydrostatic stellar atmosphere. This almost complete suppression of {velocity fluctuations} in the strong horizontal field cases indicates that magnetic tension in these configurations is able to dominate the gas dynamics, suppressing the instabilities that would otherwise lead to large velocity dispersions. 

\begin{figure*}
   \centering
    \includegraphics[scale=0.60]{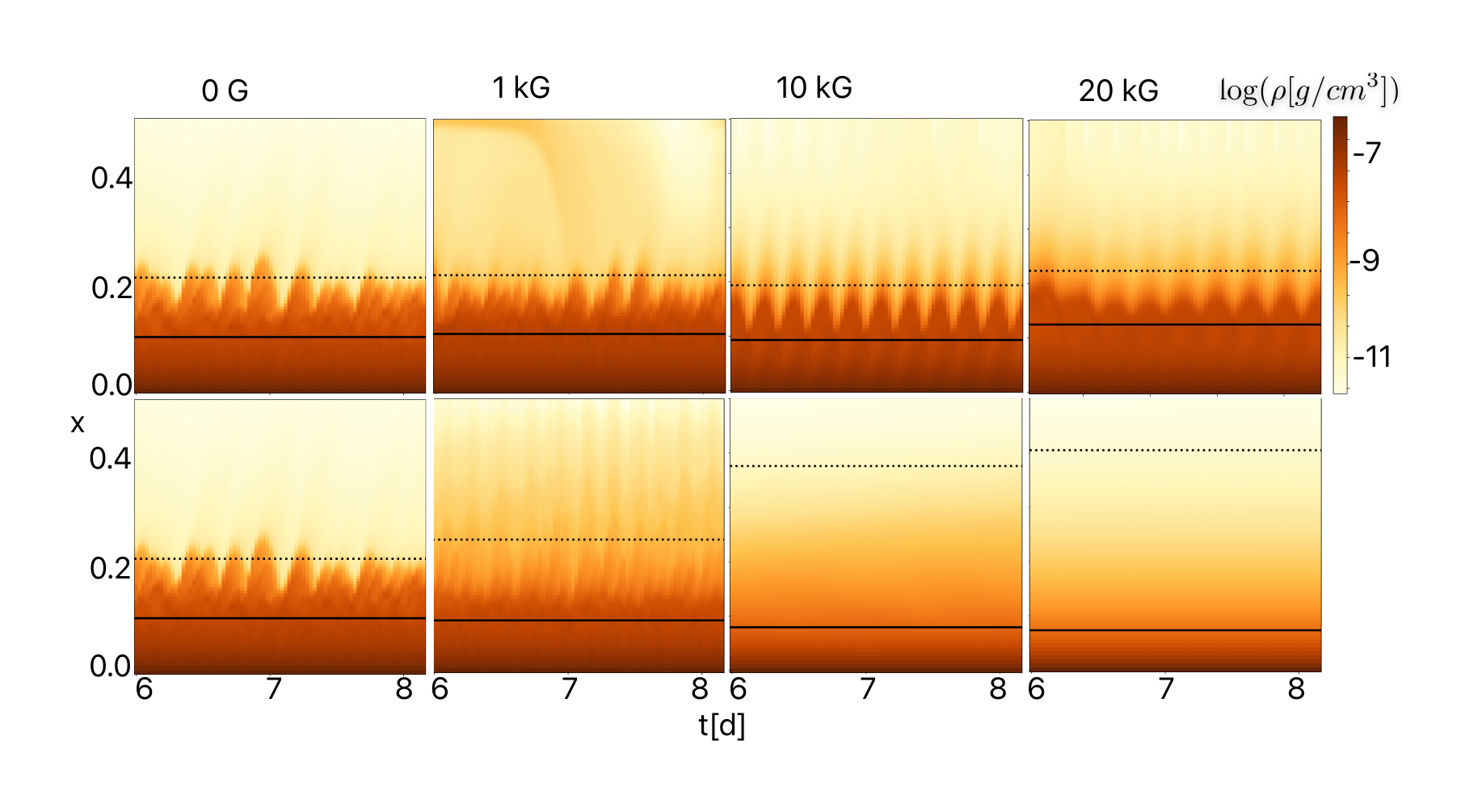}
   \caption{
Horizontally averaged density profiles are presented as a function of time, using the modified radial coordinate $x$ ($\equiv r - R_0/r$). The leftmost column serves as a non-magnetic reference. The top row displays models with radial magnetic fields of 1, 10, and 20 kG, while the bottom row showcases models with horizontal magnetic fields of the same strengths. The dotted black line indicates the photospheric radius ($r(\tau=2/3)$), and the solid black line marks the location of the iron opacity peak ($T\sim 1.6e5$ K).}  
              \label{Fig-time-density}%
    \end{figure*}

Figure \ref{Fig-time-density} illustrates the horizontally averaged density, plotted on a logarithmic scale, for all models as a function of the modified radial coordinate ($x \equiv 1 - R_0 / r$) over time. This representation emphasizes the subsurface layers, which are the primary focus of this study. The dashed black line marks the location of the average photosphere ($R_{phot}$), while the solid black line indicates the approximate position of the iron opacity peak, estimated at a temperature of $T = 1.6 \times 10^5$ K. Except for the model with the strongest horizontal magnetic field, all simulations show significant fluctuations between the iron opacity peak and $R_{phot}$. 
For reference, the leftmost column depicts the non-magnetic model, providing a baseline for comparison with the magnetic field models. 


\begin{figure*}
   \centering
   \includegraphics[scale=0.60]{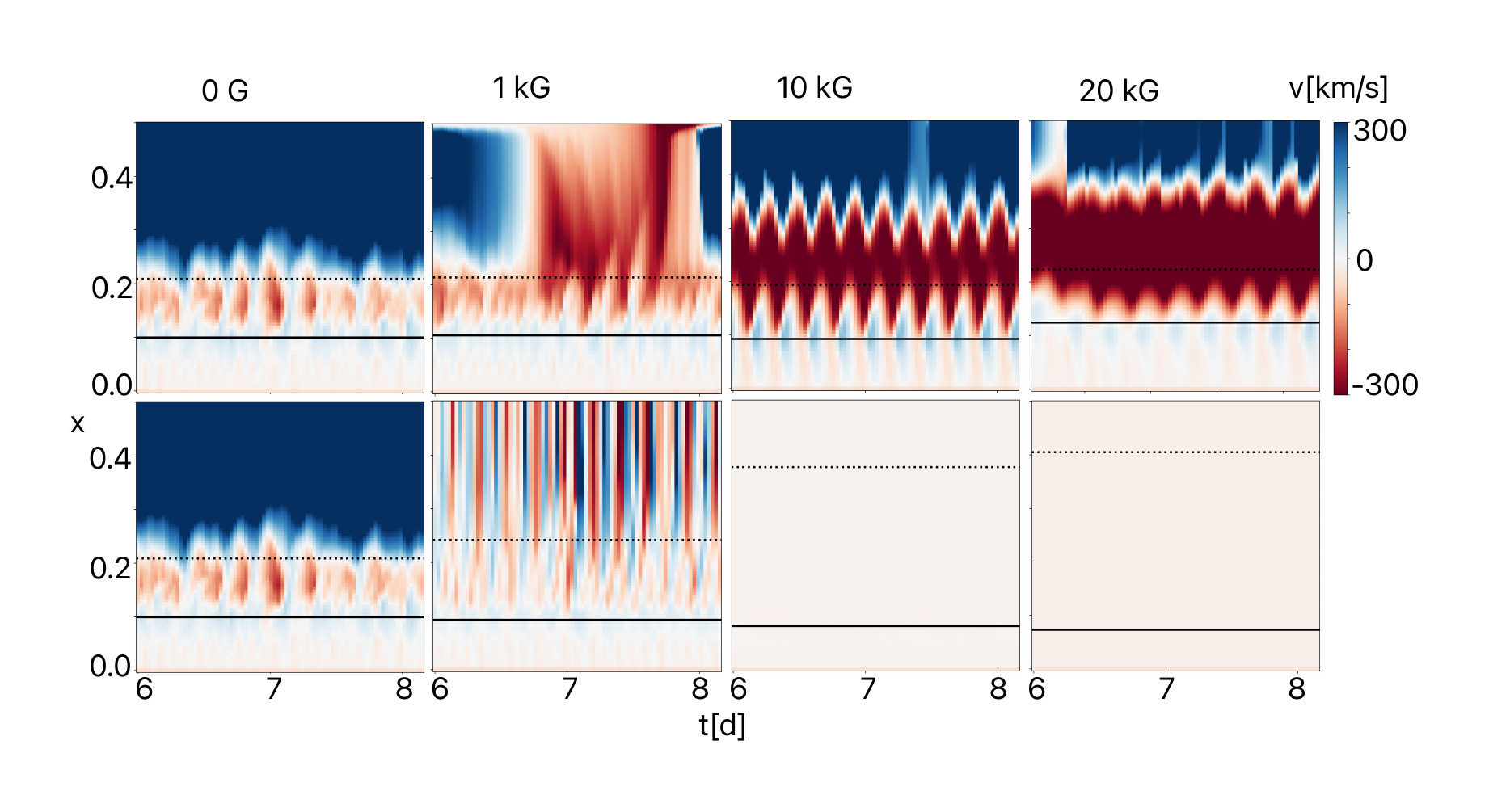}
   \caption{horizontally averaged radial velocity $\varv_r$ plotted as modified radial coordinate $x$ ($\equiv r - R_0/r$) versus time. Leftmost column represents the non-magnetic case for a direct comparison. The top row shows models with radial magnetic field of 1, 10 and 20 kG. Similarly, the bottom panels show  models with horizontal magnetic field of 1, 10 and 20 kG. Dotted black line represents photospheric radius ($R_{phot} \equiv r(\tau=2/3)$ and solid black line denotes the location of iron opacity peak ($T\sim 1.6\times 10^5$ K). Note that radial magnetic field does not inhibit up and down piston-like motions of the gas, whereas strong horizontal magnetic field can fully suppress it. This figure also demonstrates clearly how velocity dispersion is initiated near the iron opacity peak (solid black line). 
   We show only the final $\sim 3$ days of simulation for all models. Coincidentally, the model with 1 kG radial magnetic field was caught in one of the episodes of outbursts which is also visible in Figure \ref{Fig-time-density} albeit less prominently. 
   }
              \label{Fig-time-velocity}%
    \end{figure*}

Figure \ref{Fig-time-velocity} is analogous to Figure \ref{Fig-time-density}, but it depicts the radial velocity instead. 
Velocity fluctuations in all models clearly originate in the iron opacity bump region, as expected \citep{Jiang_2015, debnath_2024}. Notably, in models with a radial magnetic field orientation, the stellar surface undergoes piston-like oscillations. In contrast, models with a strong horizontal magnetic field configuration suppress these motions, preventing such oscillatory behavior. Our results demonstrate that on timescales of a few hours, radial magnetic fields do not effectively inhibit the up-and-down piston-like motions observed in the gas, mirroring the oscillatory behaviour also seen in the simulations (with weaker fields) by \citet{Jiang_2017}. Conversely, strong horizontal magnetic fields can effectively suppress these oscillations.

Table \ref{table:Results} summarizes the basic emergent stellar parameters for our models, including the stellar radius $R_{phot} \equiv r(\tau=2/3)$, effective temperature $T_{eff}$, and $T_0$, the average  temperature where magnetic and gas pressures are equal. Notably, models with moderately strong magnetic fields (both radial and horizontal) exhibit properties similar to the non-magnetic case. We do note a small deflation of the stellar envelope (i.e., a slightly smaller stellar radius) for the radial-field 1kG case as compared to the 0G run (consistent with \citealt{Jiang_2017}). This stands in sharp contrast to the significant envelope inflation found for the runs with strong horizontal magnetic fields; this is further discussed in Sect. 4.

\subsection{Suppression of motions in models with varying magnetic field strength and geometry}

\begin{figure*}
   \includegraphics[scale=0.35]{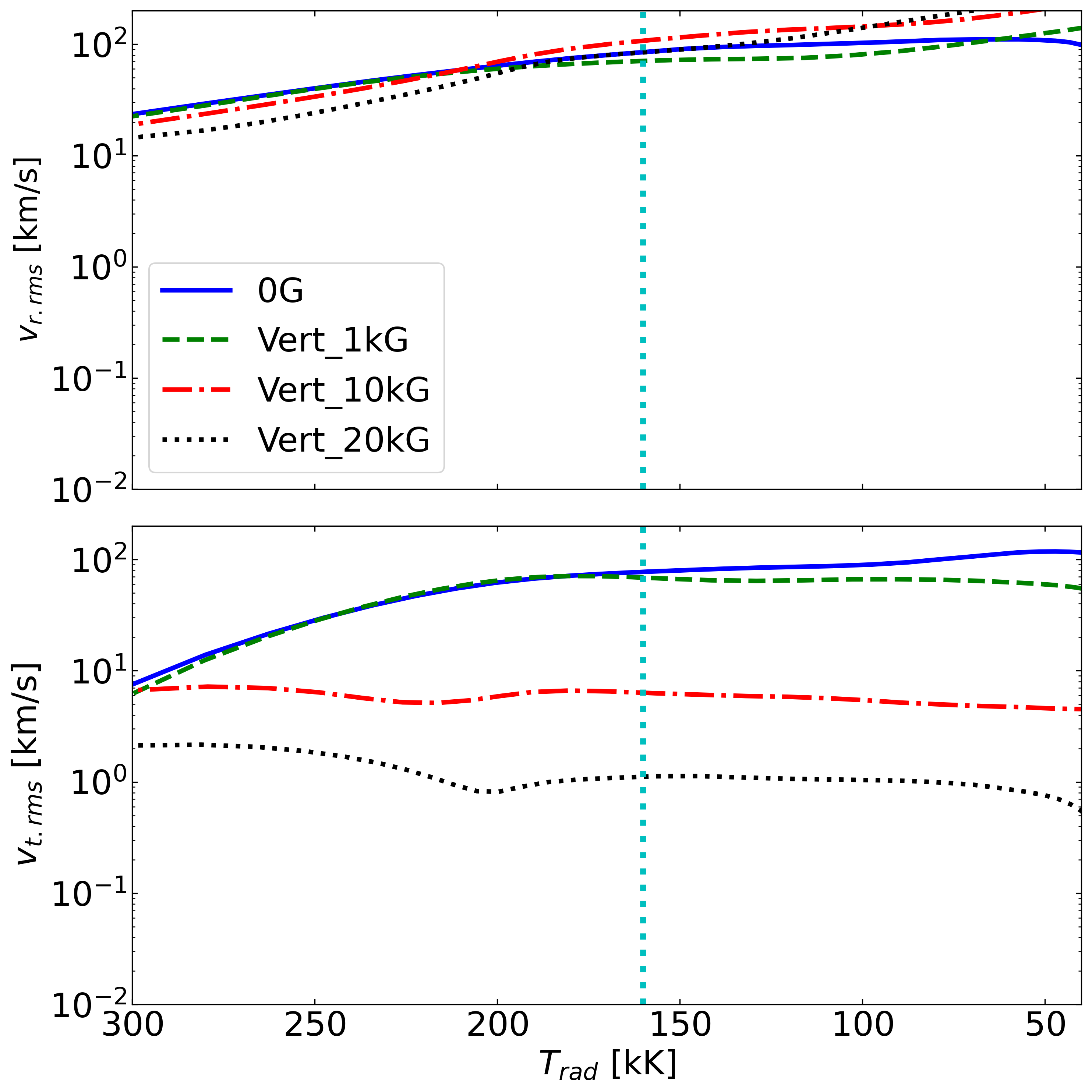}
    \includegraphics[scale=0.35]{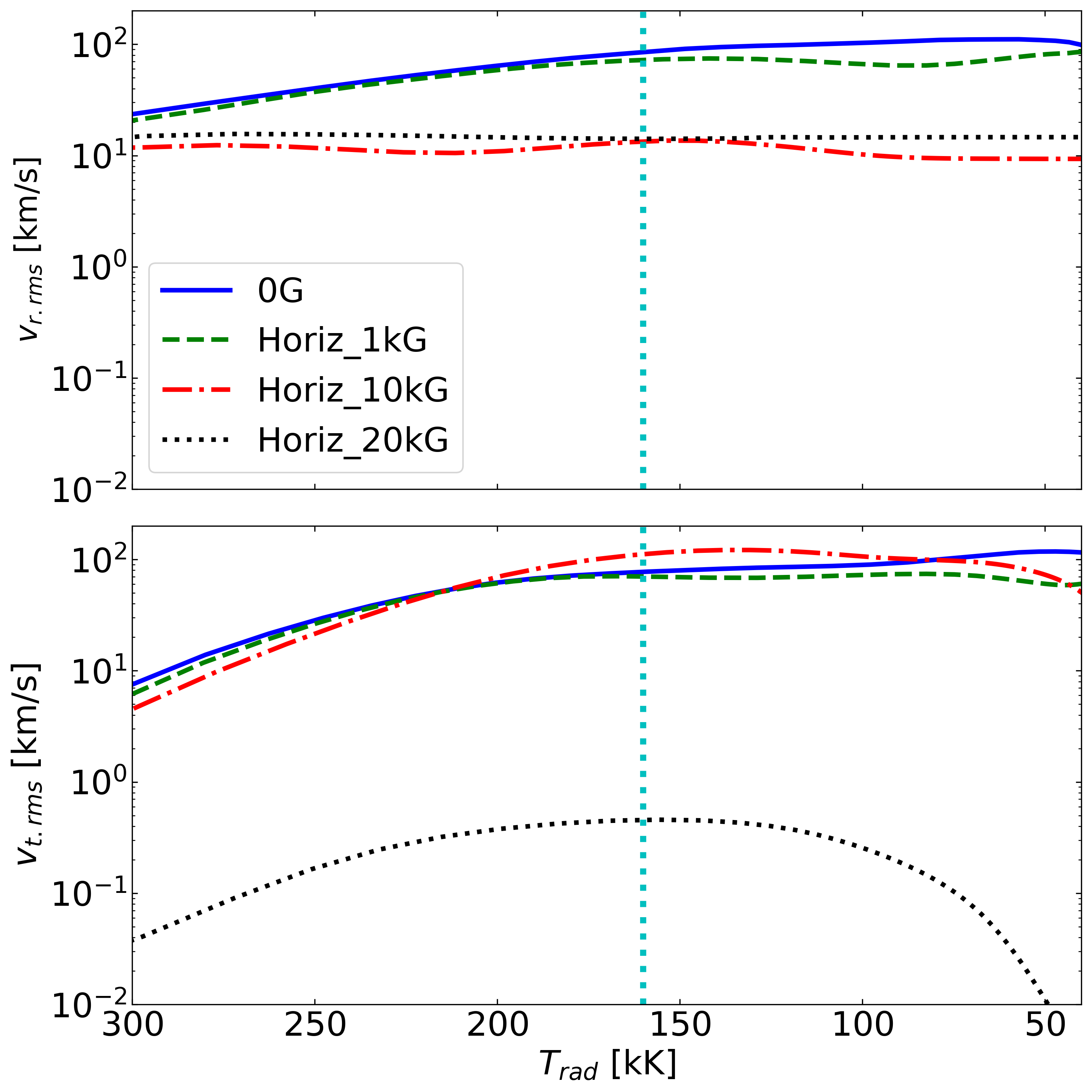}
   \caption{The root mean square (rms) values of radial (top panels) and tangential velocity perturbations (bottom panels) are shown for selected magnetic models as a function of radiation temperature, focusing on the (sub-)surface atmospheric layers before the onset of the stellar wind. The left column illustrates results for a radial magnetic field configuration, while the right column corresponds to a horizontal magnetic field configuration. It is evident that a strong radial magnetic field suppresses horizontal gas motion, while a strong horizontal magnetic field inhibits radial gas movement. Notably, in the strongest field cases (20 kG), the tangential velocity perturbations approach or fall below 1 km/s. 
   }
              \label{Fig-horiz}
    \end{figure*}

To further illustrate the effects of magnetic field strength and orientation, Figure \ref{Fig-horiz} presents the root mean square (RMS) of radial (top panels) and tangential (bottom panels) velocity perturbations ($\varv_{\rm rms}$) for the different simulation models as a very simple proxy for the strength of {photospheric velocity fluctuations}. These time-averaged RMS values were computed over approximately the final three days of the simulation to ensure that the results are not influenced by transients from the initial conditions, thereby reflecting the quasi-steady or quasi-periodic state of the atmosphere. 

In the left column, the data correspond to simulations with a radial magnetic field orientation, while the right column depicts results for models with a horizontal magnetic field orientation. The differences in RMS velocities highlight the strong dependence of {$\varv_{\rm rms}$} on the magnetic field strength and configuration. 
It is apparent from these panels that, in general, the models with strong radial magnetic fields (dotted and dot-dashed lines) demonstrate a clear suppression of horizontal gas motion, while the models with strong horizontal magnetic fields (right column, top panel) inhibit radial gas movement. 

Notably, for models with radial magnetic fields, the RMS radial velocity perturbations ($\varv_{\rm r, rms}$) remain around 100 km/s across all cases, regardless of magnetic field strength. But the situation is different for the RMS of tangential velocity perturbations ($\varv_{\rm t, rms}$). The model with a weaker 1 kG radial magnetic field (green dashed line) fails to suppress turbulence in the region of the iron opacity peak, leading to elevated tangential velocity perturbations ($\varv_{\rm t, rms} \approx 100$ km/s), which are comparable to the values seen in the non-magnetic reference model (blue line). However, for the stronger, 10 kG case, $\varv_{\rm t, rms}$ decreases to $\sim 10$ km/s while for  the strongest radial magnetic field case (20 kG), it falls even further, approaching or even falling below 1 km/s. 

In contrast, the situation reverses for models with horizontal magnetic field orientation (right column). In these models, radial gas motion is largely unaffected in the 1 kG case, where $\varv_{\rm r, rms}$ hovers around 100 km/s. However, for stronger magnetic fields ($>10$ kG), $\varv_{\rm r, rms}$ is reduced to values below 10 km/s. As for tangential velocity perturbations, all models except the strongest magnetic field case exhibit $\varv_{\rm t, rms}$ values near 100 km/s. It is only the strongest horizontal magnetic field model (20 kG) that successfully suppresses significant {velocity fluctuations} in both radial and horizontal directions, marking a significant distinction from other models.

      \begin{figure}

   \includegraphics[scale=0.35]{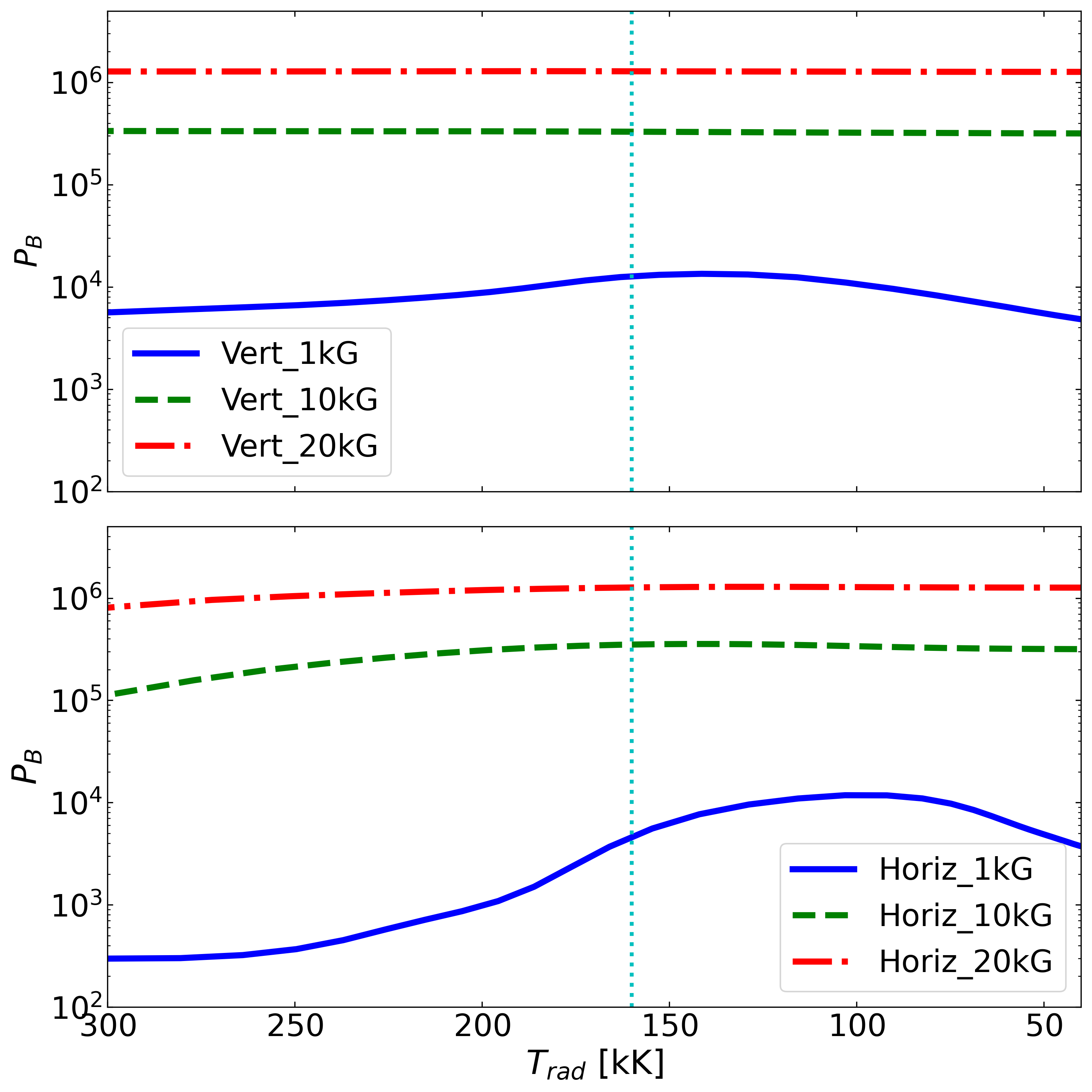}
   \caption{horizontally and time-averaged (over the final 3 days of simulation) magnetic pressure (\rm{cgs} unit)  for models with radial (top) and horizontal (bottom) magnetic field plotted as a function of radiation temperature. Note nearly the constant pressure for the strong magnetic field models whereas moderately strong models experience decompression near the base and compression around the iron opacity bump, mimicking piston-like dynamics for our numerical models.}
              \label{Fig-pB}
    \end{figure}

      \begin{figure}
   \includegraphics[scale=0.35]{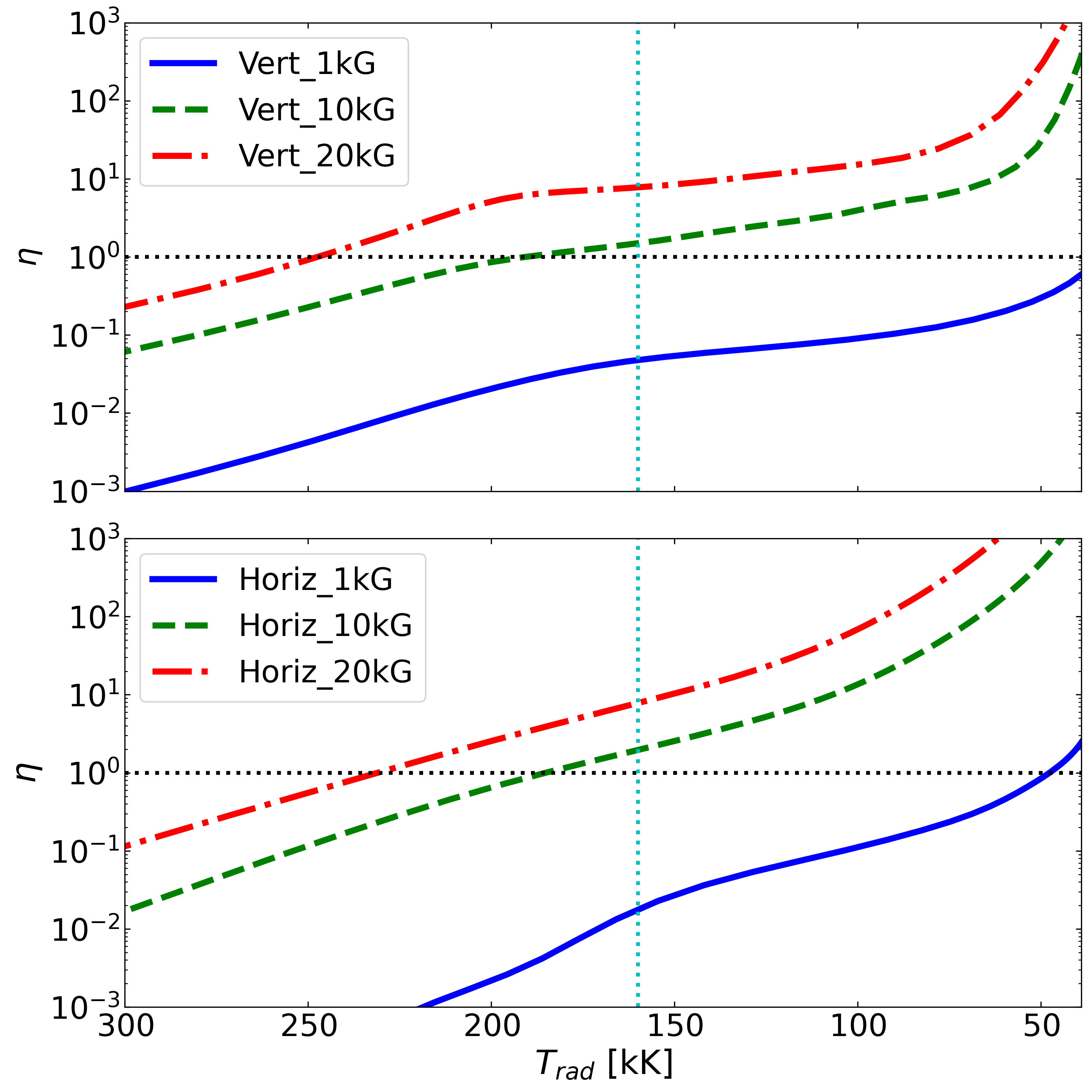}
   \caption{horizontally and time-averaged (over the final 3 days of simulation) ratio of magnetic to gas pressure ($\eta\equiv P_B/p_{gas}$) for models with radial (top) and horizontal (bottom) magnetic field plotted as a function of radiation temperature. The range has been adjusted to highlight the subsurface region near the iron opacity bump (dotted vertical line) extending to the photosphere where a stellar wind is launched.}
              \label{Fig-eta}%
    \end{figure}

  \begin{figure}
   \includegraphics[scale=0.35]{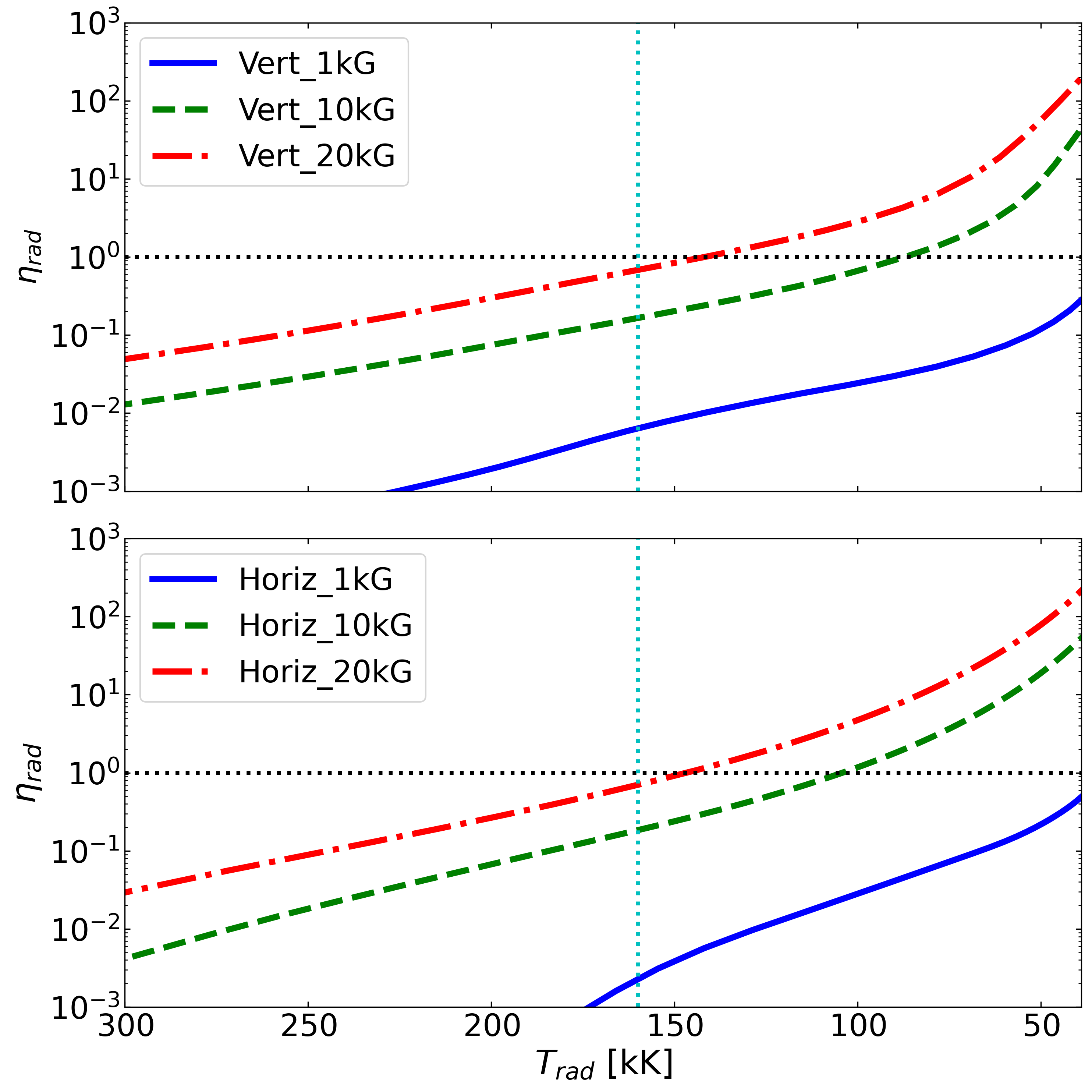}
   \caption{The same as Figure \ref{Fig-eta} but now we include radiation pressure in the definition $\eta_{rad} \equiv P_B/(p_{gas}+p_{rad})$. It shows that $\eta_{rad}$ remains below unity near iron opacity peak even for the strongest magnetic case. Yet these models suppress {atmospheric velocity fluctuations}, suggesting $p_{gas}$ plays the key role in this phenomenon.}
              \label{Fig-etarad}%
    \end{figure}

\section{Discussion} 
\label{Discussion}
In the previous section we illustrated how strong magnetic fields of order $\sim$ 10-20 kG were able to suppress the vigorous turbulent motions
generated beneath the surface in O stars (at least in the direction perpendicular to the introduced field); by contrast, fields of strengths $\sim$ 1 kG only marginally affected the turbulent motions observed in the simulations, rendering results that are qualitatively similar to those seen in non-magnetic models. These results overall agree well with observations of magnetic O-stars, where large `macroturbulence' is needed to fit optical absorption lines for all magnetic O-stars except NGC 1624-2, lending support to the hypothesis by \citet{sundqvist_2013} (see discussion in Sect. 1). 

To further analyze these simulation outcomes, Figure \ref{Fig-pB} presents the magnetic pressure, averaged horizontally and over the final three days of the simulation, for models with radial (top panel) and horizontal (bottom panel) magnetic field orientations, plotted as a function of the (horizontal and time averaged) radiation temperature (which closely aligns with the gas temperature in these optically thick regions). As expected, in the strongest magnetic field cases, the magnetic pressure remains nearly constant throughout the computational domain, reflecting the stabilizing influence of a strong field. In contrast, the weaker field case (1 kG) shows decompression near the base and compression near the iron opacity peak, indicative of the piston-like dynamics discussed above. 

Furthermore, Figure \ref{Fig-eta} shows the horizontally and time-averaged ratio of magnetic pressure to gas pressure, defined as:  
\begin{equation} 
    \eta \equiv \frac{P_B}{P_g} = \beta^{-1} 
\end{equation}
with standard 'plasma $\beta$'. For $\eta > 1$, the field thus dominates and gas parcels should be effectively forced to move along the field lines. Vice versa, for $\eta < 1$ the gas dominates and field lines should essentially be dragged along with any motions initiated in the atmosphere. We use $\eta$ here, rather than its inverse $\beta$, in order to be consistent with previous models of magnetic line-driven winds, where the definition of $\eta$ has included also the kinetic energy of the gas which is much higher than the gas pressure in line-driven outflows \citep{ud_doula_2002}. For the sub-surface and photospheric regions in focus here, however, the gas pressure is  typically higher than this kinetic energy component, especially for strongly magnetic models where the magnetic field can suppress bulk plasma motions significantly. 

\begin{table*}
\caption{The horizontally time-averaged values of $R_{phot}$, $T_{eff}$, log(g) and $T_0$— estimated temperature where magnetic and gas pressure are equal to each other—are presented for all models in this study. It is notable that models initialized with a uniform radial magnetic field exhibit properties similar to the non-magnetic case, while those with a sufficiently strong horizontal magnetic field show significant deviations from the non-magnetic scenario.
}       
\label{table:Results}      
\centering          
\begin{tabular}{c | c c c c c}  
\hline\hline       
Model & $R_{phot}$ [$R_0$] & $T_{eff}$ [kK] &  log(g) [cm s$^{-2}$]&$T_0$ [kK] \\ 
\hline  
   0G        & 1.24  & 39.98 & 3.75 & - \\
   Radial 1kG & 1.22  & 40.37 & 3.76 &61.69\\
   Radial 10kG & 1.22  & 40.05 & 3.76 &193.56\\
   Radial 20kG & 1.25  & 40.39 & 3.75 & 278.66\\
   horizontal 1kG & 1.30  & 38.95 & 3.71 &61.41\\
   horizontal 10kG & 1.49  & 36.86 & 3.59 &196.66\\
   horizontal 20kG & 1.49  & 37.97 & 3.59 & 286.43\\
\hline                  
\end{tabular}
\end{table*}

Figure \ref{Fig-eta} then shows explicitly how in the 1 kG simulation $\eta < 1$ for the whole sub-photospheric atmosphere, meaning that any gas motions initiated in these regions should be only marginally affected by the presence of the field. We note, however, that also for this case $\eta$ increases to above unity in the line-driven wind parts above the photosphere, in agreement with previous magnetic line-driven wind simulations starting from a static and smooth stellar surface \citep{ud_doula_2002}. For the 10 and 20 kG simulations, on the other hand, $\eta$ remains above unity also well below the visible stellar surface. From the figure, we can see that $\eta$ crosses unity at $T \approx 170-190$ kK for the 10 kG cases and at $T \approx 230-250$ kK for the 20 kG cases. Since atmospheric turbulence is generated in the iron opacity bump layers around $T \sim 150-200$ kK, this gives a direct explanation to the low RMS atmospheric velocities found in the previous section for the strong field cases. 

Figure \ref{Fig-etarad} next shows the horizontal- and time-averaged ratio of magnetic pressure to the total pressure that now includes radiation pressure (i.e. $\eta_{rad} \equiv P_B/(p_{gas}+p_{rad})$) as is done in \citet{MacDonald_2019}. Clearly, $\eta_{rad} < 1$ near the iron opacity peak for even the strongest magnetic field cases, yet they fully suppress the atmospheric {velocity fluctuations}. This suggests that indeed the interplay between the gas pressure and magnetic pressure is the principal factor in the suppression of turbulent motions here. 

\subsection{Analytic scaling relation for magnetic suppression of atmospheric turbulence}

We can further understand these results through the approximate scaling relation provided by \citet{sundqvist_2013}. There, the temperature $T_0$ at which gas pressure equals magnetic pressure, i.e. the temperature at which $\eta = 1$, was estimated using a simple grey atmosphere in hydrostatic equilibrium, yielding (their Eqn. 3) 
\begin{equation} 
    T_0 \approx 0.42 T_{\rm eff} B^{1/2} (\bar{\kappa}/g_{\rm eff})^{1/4} 
    \label{Eq:T0}
\end{equation} 
where $\bar{\kappa}$ is an average opacity for the sub-surface atmosphere, $g_{\rm eff}$ the effective gravity (reduced by radiative acceleration), and it has been assumed that the optical depth $\tau \gg 1$ at $T_0$.\footnote{We note here that the original scaling used the actual stellar gravity rather than one reduced by radiative acceleration (see also comments by \citealt{MacDonald_2019}), but in practise this does not change the overall scaling results because of the relatively weak $g^{-1/4}$ dependence.} Assuming opacity dominated by electron 
scattering (approximately true below the critical iron opacity region), for $T_{\rm eff}=$ 40 kK and $g_{\rm eff} = 4050$ we obtain $T_0 \approx 50$ kK for $B=$ 1 kG, $T_0 \approx 160$ kK for $B=$ 10 kG, and $T_0 \approx 230$ kK for $B =$ 20 kG. These characteristic analytic estimates compare reasonably well with the detailed simulation results above, {as do similar estimates of critical field strengths above which the magnetic field can suppress turbulence by \citet{MacDonald_2019} and \citet{Jermyn_2020}. In those studies, however, the kinetic energy of the turbulent gas was estimated by simple mixing-length-theory (MLT) for convection, whereas our O-star simulations (here and in \citealt{debnath_2024}) show that such MLT cannot be used to estimate turbulent velocities arising from instabilities in the sub-surface iron-bump region of O-stars. As such, eqn. \ref{Eq:T0} above provides a straightforward, simple rationale} for understanding how potential suppression of photospheric turbulence by the magnetic field scales with basic atmospheric parameters.

\subsection{Magnetically induced atmospheric inflation of luminous massive stars} 

{As shown in Table \ref{table:Results} and Figure \ref{Fig-time-density}, the size of the photosphere, \( R_{phot} \), depends not only on the magnetic field strength but also strongly on its orientation. For models with a radial magnetic field, \( R_{phot} \) is fluctuating (because of the oscillatory motions discussed above) but its average value remains only marginally affected, regardless of the field strength. However, for models with a strong horizontal magnetic field, a significant increase in \( R_{phot} \) is observed in the simulations. 

This inflation effect very likely occurs in the models precisely \textit{because} turbulent motions are suppressed; indeed, it mimics the inflationary behaviour seen in 1D models of luminous massive stars assuming a radiative stellar envelope in hydrostatic equilibrium (e.g., see discussions in \citealt{petrovic_2006, grafener_2012, luka_2021,debnath_2024}); the key physics driving this mechanism is the higher opacity in the iron-bump region, which pushes the luminous star toward its effective Eddington limit. The only way the envelope then can remain in hydrostatic balance is by expansion of its envelope and atmosphere. Since radiative luminosity is conserved, the larger radius then also leads to a lower stellar effective temperature, as seen in  Table \ref{table:Results}. We note that this inflation effect thus generally becomes more prominent the closer the radiative massive star is to the classical Eddington limit (see also discussion in \citealt{owocki_2015}). 

This tentatively suggests that the observable surface of luminous massive stars with strong global dipole fields could be significantly asymmetric and deformed, where an observer viewing from the magnetic equator might see a significantly larger and less variable star than an observer viewing from above the magnetic pole. It is important to note, however, that our local 2D simulations neglect potentially important global transport effects. For example, if rapid lateral diffusion of photons would be able to equilibrate polar and equatorial regions, this could significantly reduce the pronounced latitudinal effects on stellar temperature and radius found here.

\subsection{Inefficient convective instability in iron opacity peak region of O-stars}

The very turbulent O-star atmospheres seen in the RHD simulations by \citet{Jiang_2015} and \citet{debnath_2024} normally have been interpreted in terms of sub-surface convection arising in the iron opacity peak region. However, as mentioned above (see also discussions in \citealt{Jiang_2015} and \citealt{debnath_2024}) energy transport by enthalpy (that is, by convection) is actually very inefficient in this region. Indeed, detailed linear stability analysis (\citealt{bs03}, van der Sijpt et al, in prep.) demonstrates that more instabilities are present in the region (e.g., radiatively driven unstable sound waves and classical Rayleigh-Taylor), and that structure growth may rather be dominated by these. This is supported by the simulations here, especially the strong 20 kG vertical field case. Namely, 
linear stability analysis including magnetic fields \citep{bs03} shows that a strong vertical field that suppresses horizontal motion will also be able to completely suppress convection (simply since there is no gravity mode instability in the vertical direction). A similar reasoning was used also in \citet{MacDonald_2019} to argue that a strong vertical field should stabilise the atmosphere against convection while an equally strong horizontal field should not (see their Sect. 2.2). 

This is in stark contrast to our 2D RMHD model results, where large radial motions still occur in the simulations with a strong vertical field; that is, the atmospheric motions in those simulations cannot stem from the convective instability. Rather they very likely arise because the effective Eddington limit is exceeded already at the sub-surface iron opacity peak, launching 'failed wind outflows' that stagnate and start to fall back as the opacity decreases when approaching the optical surface. Since the outwards suction effect from line-driving is efficient only near and above this surface, and the gas is prevented by the magnetic field to move horizontally, the outcome is a radially oscillating atmosphere characterised by alternating infalling and outflowing regions. By contrast, in the simulations with a strong horizontal field, such radial motion is effectively suppressed, leading instead to a quasi-hydrostatic atmosphere and thereby to the inflation effect discussed above. Since curvature effects will break the symmetries of the local 2D simulations presented here, it is likely that (if real) these radial atmospheric motions would be constrained to regions close to the magnetic pole.

\section{Summary and future work}
\label{summary}

The work here has provided new insights into the role of strong magnetic fields in suppressing velocity fluctuations within the sub-surface layers of O-stars, particularly in those with magnetic fields exceeding 10 kG in horizontal direction. Our 2D RMHD simulations reveal that such strong magnetic fields can effectively suppress the large turbulent motions typically associated with the atmospheres of these stars. This finding offers a plausible explanation for the unique case of NGC 1624-2, which possesses a very strong $\sim$ 20 kG dipolar field and does not exhibit the additional line-broadening (`macroturbulence') observed in optical absorption lines of other magnetic and non-magnetic O-stars \citep{sundqvist_2013}. Analogously, this suppression also means that if significant stochastic low-frequency variability \citep[e.g.,][]{Bowman24} were to be found in NGC 1624-2, that would speak against an origin of such low-frequency variability in the sub-surface turbulent region of massive stars (see also discussion in \citealt{Jermyn_2020}).

In models with radial magnetic field orientation, the stellar surface exhibits piston-like oscillations, a signature of vertical motion induced by the field-aligned configuration. These oscillations are characterized by periodic compression and expansion of the gas, which is able to flow freely along the radial magnetic field lines. The motion occurs on a timescale of about half a day, quite similar to the oscillatory behaviour found in the simulations by \citet{Jiang_2017} for cases with weaker fields up to 1kG. {As discussed above, since there is no gravity mode instability in the radial direction this also shows that it is not convection that primarily drives these atmospheric motions.}

On the other hand, in models with a strong horizontal magnetic field, these piston-like motions are significantly suppressed. The horizontal field orientation effectively inhibits the radial movement of gas, likely due to the magnetic tension forces acting to resist vertical displacement. Instead of oscillating, the gas remains relatively stationary, suggesting that the magnetic field in this configuration is able to fully constrain the large velocity fluctuations that would otherwise occur. This stark difference between the radial and horizontal magnetic configurations highlights the importance of magnetic field geometry in governing the dynamics of the atmospheres of magnetic massive stars. Strong horizontal fields, in particular, appear to almost completely quench the instabilities driving turbulence in massive-star envelopes, leading to a more stable atmosphere.

As discussed in the previous section, this quenching also leads to significant envelope inflation in the simulations with strong horizontal magnetic fields. By contrast, an equally strong radial field shows a fluctuating photosphere but with an average value that is only marginally different from the non-magnetic simulations. This may suggest that in luminous massive stars with strong global dipole fields, the observable surface above the magnetic equator might be significantly larger, cooler, and less variable than the surface above the magnetic pole.

While these results are promising and seem to offer a natural explanation for the broad characteristics of observed optical absorption lines in magnetic O-stars, it is important to keep in mind that they are based on 2D local simulations, which inherently limit the ability to capture the full complexity of the stellar atmospheres. Full, global 3D RMHD simulation including curvature effects will indeed be needed to confirm the presence of the above-mentioned radial piston-like motions above the pole as well as the potential atmospheric inflation above the magnetic equator discussed above. In particular, rapid lateral photon diffusion may  induce a more uniform 
thermal structure in latitude than found in the local simulations here and might also affect the strength of vertical motions in polar regions. Future work should thus focus on extending these simulations to full 3D global models including curvature effects and the spherical divergence of the magnetic field. Such an approach would allow for a more consistent and systematic investigation of atmospheric turbulence suppression. This would also be required for a more quantitative analysis of the narrow absorption lines observed for NGC 1624-2 
\citep{Wade_2012b, DavidUraz_2021}, since observer's viewing angle and global magnetic field topology both will affect the surface-integrated spectral line formation. 

In conclusion, our work here provides a foundation for such future studies aiming to further explore the intricate interplay between magnetic fields and atmospheric turbulence in O-stars. Such studies are essential for a deeper understanding of the variability observed in the spectra of magnetic massive stars and the role of magnetic fields in shaping their atmospheres.

\begin{acknowledgements}
We are grateful to Stan Owocki for providing fruitful suggestions in performing this study.

A.u.-D. acknowledges NASA ATP grant number 80NSSC22K0628 and support by NASA through Chandra Award number TM4-25001A issued by the Chandra X-ray Observatory 27 Center, which is operated by the Smithsonian Astrophysical Observatory for and on behalf of NASA under contract NAS8-03060.

The computational resources used for this work were provided by the Penn State University Collab Roar system. DD, NM, and JS acknowledge the support of the European Research Council (ERC) Horizon Europe under grant agreement number 101044048 (ERC-2021-COG, SUPERSTARS-3D). JS and A.u.-D. acknowledge support from the Belgian Research Foundation Flanders (FWO) Odysseus program under grant number G0H9218N. JS further acknowledges support from FWO grant G077822N. RK is supported by FWO projects G0B4521N and G0B9923N and received funding from the European Research Council (ERC) under the European Union Horizon 2020 research and innovation program (grant agreement No. 833251 PROMINENT ERC-ADG 2018). We gratefully thank our referee, Matteo Cantiello, for useful suggestions that helped us add some important nuances to the discussion of our results. Finally, the authors would like to thank all members of the KUL-EQUATION group for fruitful discussion, comments, and suggestions We made significant use of the following packages to analyze our data: {\fontfamily{qcr}\selectfont NumPy} \citep{harris_2020}, {\fontfamily{qcr}\selectfont SciPy} \citep{virtanen_2020}, {\fontfamily{qcr}\selectfont matplotlib} \citep{hunter_2007}, {\fontfamily{qcr}\selectfont Python amrvac\_reader} \citep{keppens_2020}.

\end{acknowledgements}

\bibliographystyle{aa}
\bibliography{references_ostar} 

\end{document}